\documentclass[prl,aps,twocolumn,superscriptaddress,showpacs]{revtex4-1}

\usepackage{dcolumn}
\usepackage{amsfonts, amssymb, amsmath}
\usepackage{bm}
\usepackage{epsf,epsfig}
\usepackage{graphics,psfrag}
\usepackage{graphicx,psfrag}
\usepackage{wasysym}
\usepackage{natbib}

\usepackage[utf8x]{inputenc}
\usepackage[T1]{fontenc}
\usepackage{ulem}

\usepackage[usenames,dvipsnames]{xcolor}

\usepackage{color}
\definecolor{darkblue}{rgb}{0,0,0.6}
\definecolor{darkred}{rgb}{0.6,0,0}

\usepackage{hyperref}
\hypersetup{
bookmarksopen=true,
pdftitle="Nonlinear Rheology in a Model Biological Tissue",
pdfauthor="", 
pdftoolbar=false,			
pdfstartview={FitH},		
pdfmenubar=true,			
pdfhighlight=/O,			
colorlinks=true,			
urlcolor=darkblue,
citecolor=darkblue,		
linkcolor=MidnightBlue,	
}

\newcommand{\argp}[1]{\left(#1\right)}
\newcommand{\valabs}[1]{\vert #1\vert}
\newcommand{\moy}[1]{\left\langle  #1 \right\rangle }

\definecolor{darkgreen}{rgb}{0.0, 0.5, 0.0}

\begin{document}

\title{
Nonlinear Rheology in a Model Biological Tissue
}

\author{D.~A. Matoz-Fernandez}
\email[]{daniel-alejandro.matoz-fernandez@univ-grenoble-alpes.fr}
\affiliation{Universit\'e Grenoble Alpes \& CNRS, LIPHY, F-38000 Grenoble, France}
\author{Elisabeth Agoritsas}
\email[]{elisabeth.agoritsas@lpt.ens.fr}
\affiliation{Universit\'e Grenoble Alpes \& CNRS, LIPHY, F-38000 Grenoble, France}
\affiliation{Laboratoire de Physique Th{\'e}orique, ENS \& PSL University, UPMC \& Sorbonne Universités, F-75005 Paris, France}
\author{Jean-Louis Barrat}
\affiliation{Universit\'e Grenoble Alpes \& CNRS, LIPHY, F-38000 Grenoble, France}
\author{Eric Bertin}
\affiliation{Universit\'e Grenoble Alpes \& CNRS, LIPHY, F-38000 Grenoble, France}
\author{Kirsten Martens}
\affiliation{Universit\'e Grenoble Alpes \& CNRS, LIPHY, F-38000 Grenoble, France}

\begin{abstract}

The rheological response of dense active matter is a topic of fundamental importance for many processes in nature such as the mechanics of biological tissues.
One prominent way to probe mechanical properties of tissues is to study their response to externally applied forces.
Using a particle-based model featuring random apoptosis and environment-dependent division rates,
we evidence a crossover from linear flow to a shear-thinning regime with an increasing shear rate.
To rationalize this nonlinear flow we derive a theoretical mean-field scenario that accounts for the interplay of mechanical and active noise in local stresses.
These noises are, respectively, generated by the elastic response of the cell matrix to cell rearrangements and by the internal activity.

\end{abstract}


\maketitle




Mechanical stimuli on single cells \cite{Moeendarbary2014} and cell assemblies \cite{lecuit2007cell} play an important role in biology, for example in the mechanics of biofilms \cite{Billings2015} as well as for medical issues \cite{jacobs2013introduction, vandenBerg2015}. Furthermore, mechanical sensing has been shown to be of vital importance in cancer growth \cite{paszek2005tensional, delarue2014compressive, Nagelkerke2015} and morphogenesis \cite{popovic2016active, etournay2015interplay}. Driven by advances in experimental cell tracking techniques \cite{Trepat2009, angelini2011glass, etournay2016tissueminer, nnetu2013slow}, this topic has gained a lot of importance in recent years. The mechanical response of cell aggregates under deformation has been shown to exhibit elastic, elastoplastic and viscous flow behavior depending on the forces applied and the time scale of observation considered \cite{Marmottant2009, angelini2011glass, heermann2015eye}. Recently there have been many efforts to understand the origin of these different mechanical regimes. It has been shown that both self-propulsion \cite{Berthier2014, Mandal2016, Bi2016, Berthier2016}, as well as cell division and apoptosis \cite{Ranft2010,Basan2011, Matoz2016a} are processes able to fluidize a confluent cell assembly, which appears to be arrested in a glassy configuration otherwise.

In this Letter we go beyond the study of the specific fluidization 
mechanism and the corresponding linear flow regime \cite{Basan2011}. We 
investigate the flow properties of a confluent tissue under shear using 
a particle-based model that incorporates activity in the form of cell 
division and apoptosis \cite{Matoz2016a}. We find 
that the internal activity gives rise to a fluidization of the tissues at shear rates 
smaller than a time scale set by the apoptosis rate, followed by a 
shear-thinning regime, well described by a Herschel-Bulkley flow curve at higher shear rates.
These findings are in agreement with experimental studies on epithelial cell monolayers \cite{nnetu2013slow}, 
which showed that the structural relaxation time of their tissue was purely governed by the cell division time in the high density regime.

\begin{figure}[!t]
\begin{center}
\includegraphics[width=0.7\columnwidth, clip]{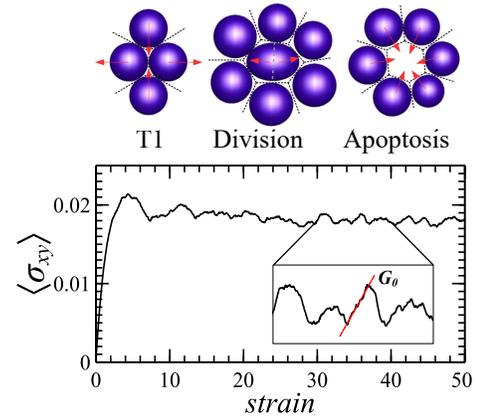}
\end{center}
\caption{(\textit{Top}) Microscopic events in a system with cell 
division and apoptosis under shear. From left to right, T1 event 
(passive rearrangement), cell division and apoptosis (active rearrangements).
(\textit{Bottom}) Typical macroscopic 
stress-strain curve, the average slope of the
increasing elastic parts on the curve corresponds to the elastic modulus $G_{0}$.
}
\label{fig:1}
\end{figure}

In analogy to the flow of soft matter, such as emulsions or foams, we 
propose a statistical description to derive an analytical prediction 
for the complex flow curve in confluent tissues. At the core of this 
description is an elastoplastic picture: the inactive cell assembly 
responds elastically to external forcing like a solid up to a threshold 
above which it is able to locally yield through cell-cell 
rearrangements leading to plastic flow as shown in the stress-strain curve in the bottom panel of Fig.~\ref{fig:1}. The local rearrangements (T1 events, see sketch in Fig.~\ref{fig:1}) lead to a long-range elastic response of the surrounding medium that will create a mechanical noise \cite{Puosi_Olivier_Martens_2015_Softmatter11_7639}. We argue that the elastic perturbation created through the internal activity, for example via cell division and apoptosis (sketch in Fig.~\ref{fig:1}), creates an additional active noise. The interplay of these different mechanisms leads to an interesting nontrivial flow behavior. 
We rationalize these findings using a mean-field description that extends 
the H\'ebraud-Lequeux model 
\cite{hebraud_lequeux_1998_PhysRevLett81_2934} of athermal yield stress fluids.

In spite of the inherent complexity of tissue mechanics, many interesting
collective phenomena at tissue level can be described using 
simple models in which cell-cell interactions are treated as soft 
interactions between particles~\cite{Drasdo2007,zimmermann2016contact}. In this spirit, we model 
a tissue as a collection of $N$ soft spherical particles with 
radii $b_{i}$ uniformly distributed in a range of $0.85$ to $1.15$. Moreover, in order to mimic the real behavior of cells in epithelial sheets we consider adhesion and excluded volume as a combination of attractive and repulsive forces~\cite{Matoz2016a,Szabo2006},
\begin{equation}
\label{eq-def-forces}
\mathbf{F}_{ij} =    
\begin{cases}
-k b_{ij} \left(1- \frac{r_{ij}}{b_{ij}}\right) \mathbf{\hat{r}}_{ij}&\mbox{if } 0 \leq \frac{r_{ij}}{b_{ij}} \leq \epsilon + 1 \vspace{0.1cm}\\
k b_{ij}  \left(2\epsilon + 1 - \frac{r_{ij}}{b_{ij}}\right) \mathbf{\hat{r}}_{ij} &\mbox{if }  \epsilon< \frac{r_{ij}}{b_{ij}}-1 \leq 2\epsilon,
\end{cases}
 \end{equation}
where $k$ is the stiffness constant, $b_{ij} = b_{i} + b_{j}$ is the sum of the particle radii and $\epsilon$ is the ratio of the maximal attractive and maximal repulsive forces.
The cell centroids $\mathbf{r}_{i}(t)$ follow an overdamped dynamics,
${\partial_t \mathbf{r}_{i}(t) =  \mu \mathbf{F}_{i}}$,
where $\mu$ is the mobility coefficient~\cite{Henkes2011}.
In addition, activity is introduced via apoptosis and cell-division rates.
Apoptosis (as well as possibly other cell death mechanisms) is included by removing cells randomly at a constant rate $a$.
On the other hand, as in real epithelial tissues, the contact inhibition process~\cite{puliafito2012} is modeled via a density-dependent division rate ${d_i = d_{0}(1 - z_i/z_{\rm max})}$,
with $d_{0}$ the division rate amplitude, $z_i$ the number 	of contact neighbors of particle $i$, and $z_{\rm max}$ the maximum number of contact neighbors allowed.
After any division, the new daughter cell is placed on top of the mother cell.
In order to prevent any numerical instability, the total force exerted by the mother and daughter cells on the surrounding cells is kept continuous, by applying only half the force immediately after cell division, and then progressively increasing the applied force to reach again a nominal force applied on each cell~\cite{Matoz2016a}.


We carried out 2D simulations of the model with fixed values ${z_{\text{max}}=6}$, ${\epsilon=0.05}$~\cite{SM}.
By setting ${\mu = k = 1}$ we set the unit time to the elastic relaxation time ${\tau_{\text{el}}=(\mu k)^{-1}}$, typically of the order of minutes \cite{Marmottant2009}.
We take as a control parameter the apoptosis rate $a$, where $a^{-1}$ is in experiments of the order of half an hour to half a day \cite{nnetu2013slow, puliafito2012collective}.
Since varying $a$ at the fixed maximal division rate $d_0$ would lead to large variations in packing fraction, we rather fix $d_0$ through the relation ${a/d_0=0.1}$ (consistent with the choice of the other parameters and typical experimental values \cite{nnetu2013slow}), leading to limited changes of the packing fraction.

\begin{figure}[t!]
\begin{center}
\includegraphics[width=0.7\columnwidth, clip]{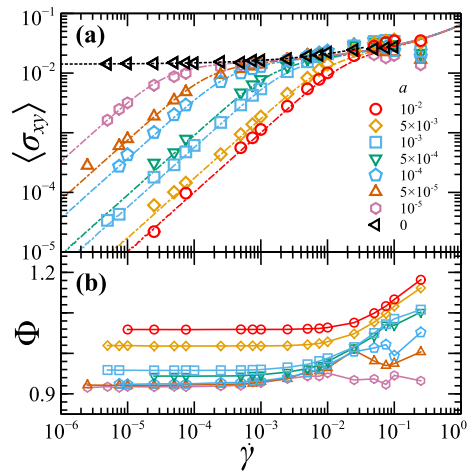}
\end{center}
\caption{{\it Flow curves for passive and active systems --}
(a)~Steady-state average shear stress $\langle\sigma_{xy}\rangle$ 
versus the applied external shear rate $\dot{\gamma}$ for different apoptosis rates $a$.
The symbols indicate the microscopic simulations results for the
different apoptosis rates.
The dashed line corresponds to a Herschel-Bulkley fit of the passive system
with packing fraction $\Phi\approx0.94$. 
The dashed-dotted lines are the mean-field model fits with fitting parameter $D_0$.
(b)~Shear-rate dependence of the corresponding packing fraction $\Phi$ for the same values of the apoptosis rates.}
\label{fig:2}
\end{figure}

After a steady state has been reached in the absence of shear, with the average division rate balancing the apoptosis rate,
we impose a constant shear rate $\dot{\gamma}$ via deformation of a triclinic box with periodic boundary conditions.
We apply a strain of more than 10 to ensure that the steady state has been reached and measure the macroscopic shear stress ${\sigma_{xy}(t)}$ as illustrated in Fig.~\ref{fig:1}.

We have established that the homeostatic properties of the system do not strongly depend on $a$~\cite{SM}. 
For low enough shear rate $\dot{\gamma}$ and apoptosis rate $a$,
the packing fraction is constant [Fig.~\ref{fig:2}(b)].
A careful analysis of the packing fraction $\Phi$ as a function of $a$ and $\dot{\gamma}$ [Fig.~\ref{fig:2}(b)] however reveals two different regimes in which the packing fraction deviates from this constant value.
First, considering a fixed low shear rate (typically ${\dot{\gamma}<10^{-3}}$),
the packing fraction $\Phi$ increases with activity if ${a \gtrsim 10^{-3}}$.
This is understood as an interplay between the division rate $d_0$ (equal to $0.1 a$) and the mechanical relaxation rate $\mu k$ in the soft repulsive potential~\cite{Matoz2016a}.
As long as the elastic relaxation time ${\tau_{\text{el}}=(\mu k)^{-1}}$ remains
small with respect to the typical time ${\tau_{\text{a}}=(d_{0})^{-1}}$ between two divisions involving the same cell, the elastic relaxation processes remain independent and the packing fraction remains constant.
When, in constrast, ${\tau_{\text{a}}\ll \tau_{\text{el}}}$, multiple divisions occur during the elastic relaxation, and the resulting packing fraction depends on the activity $a$.
The second regime is related to the effect of a strong enough shear rate, typically ${\dot{\gamma}>10^{-3}}$. In this case, we observe that a fast and large deformation of the box produces a rapid decrease of the number of contact neighbors followed by an increase of the division rate,
eventually leading to a steady state with a higher packing fraction.

\begin{figure}[t!]
\begin{center}
\includegraphics[width=0.7\columnwidth, clip]{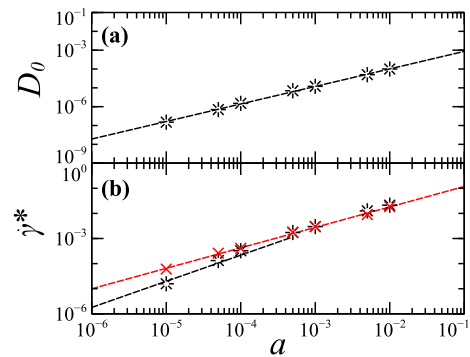}
\end{center}
\caption{
(a) Fitted value of the diffusion coefficient $D_0$ of the active noise as a function of the apoptosis rate $a$ from the mean-field fit on the simulation data. The dashed
line is a power-law fit with an exponent 0.94.
(b) Activity-dependent crossover shear rate $\dot{\gamma}^*(a)$ measured from the 
flow curves (see text for details). Crosses are determined values from the pure simulation data, whereas stars come from the mean-field fit of the simulations.
}
\label{fig:3}
\end{figure}

The corresponding flow curves, i.e., the steady-state macroscopic stress ${\langle \sigma_{xy} \rangle}$ as a function of $\dot{\gamma}$, are shown in Fig.~\ref{fig:2}(a) for different values of the apoptosis rate.
In the absence of activity (${a=0}$)
the system exhibits a nonlinear rheology as observed in foams; it is well known that this type of dynamics is then characterized at low shear rates by a Herschel-Bulkley flow curve
${\langle\sigma_{xy}\rangle = \sigma_y^{(d)} + A_{\text{HB}} \dot{\gamma}^n}$.
In other words, in the limit of zero shear rate, the macroscopic stress takes a finite value, known as the dynamical yield stress $\sigma_y^{(d)}$, and then increases with the shear rate following a power-law behavior \cite{herschel_bulkley_1926_kolloid-zeitschrift39_291}.
In our case, we obtain ${\sigma_y^{(d)}=0.014}$, ${A_{\text{HB}}=0.065}$ and 
${n\approx0.5}$
for ${\dot{\gamma} \leq 0.025}$ (see the dashed curve), this exponent  being consistent with those observed in foams 
and in recent molecular dynamics simulations \cite{Puosi_Olivier_Martens_2015_Softmatter11_7639}.
On the other hand, a finite activity $a >0$ prevents the system from  having a finite yield stress $\sigma_y^{(d)}$, 
leading  to a linear behavior at low shear rates,
with a viscosity that decreases when $a$ increases.
Here the new feature is the crossover, at a shear rate $\dot{\gamma}^*$ controlled by the activity, from a Newtonian to a Herschel-Bulkley behavior.
Defining the crossover  as the intersection between the linear regime and the 
plateau, we have plotted ${\dot{\gamma}^*(a)}$ in Fig.~\ref{fig:3}(b),
obtaining an almost linear dependence ${\dot{\gamma}^* \sim a^{0.82}}$.
We emphasize that, at least for small enough activity ($a \lesssim 10^{-4}$), the crossover from activity-driven fluidization to a yield-stress (plateau) behavior occurs at a constant packing fraction.
On the contrary, the stress increase at large shear rate $\dot{\gamma}$ 
results from both standard elastoplastic effects and the increase of the packing fraction.


The crossover from linear to nonlinear flow behaviors of the sheared active system can be captured via a minimal mean-field description that focuses on the dynamics of the local shear stress.
For this purpose, we use an athermal-local-yield-stress model \cite{agoritsas_2015_EPJE38_71,agoritsas_martens_2016_ArXiv-1602.03484}, which generalizes the original H\'ebraud-Lequeux model \cite{hebraud_lequeux_1998_PhysRevLett81_2934}.
These models usually do not take into account any active contribution to the local stress fluctuations. As shown in Fig.~\ref{fig:2}(a), this active contribution is however a key ingredient for fluidizing the system and is thus introduced explicitly thereafter.
In a simplified mean-field picture, the dynamics of the local stress can be modeled by a modified Langevin dynamics,
${\partial_t \sigma(t) = G_0 \dot{\gamma} + \xi_{\text{mec}}(t)}$
with $G_0$ the average local elastic modulus,
${\dot{\gamma}}$ the external constant shear rate,
and ${\xi_{\text{mec}}(t)}$ the mechanical noise.
In addition, once ${\sigma(t)}$ exceeds a typical threshold ${\sigma_c}$, a local plastic event randomly occurs at a fixed rate $1/\tau$, which would in turn fully relax the local stress and thus reset $\sigma(t)$ to $0$.
Here, ${\xi_{\text{mec}}(t)}$ is modeled by a Gaussian noise with zero mean.
Furthermore, at the time scale we consider, we can neglect its time correlation, i.e.,
${\moy{\xi_{\text{mec}}(t) \xi_{\text{mec}}(t')} = 2 D(t) \delta(t-t')}$,
where the brackets corresponds to average over time, and $D(t)$ is the stress diffusion coefficient.
At low shear rate, a natural way to introduce activity is to distinguish two contributions to the noise, i.e., ${\xi_{\text{mec}}(t)=\xi_{\text{pl}}(t)+\xi_{\text{act}}(t)}$.
The noise ${\xi_{\text{pl}}(t)}$ accounts for the plastic events triggered by the external driving throughout the system, and has a time-dependent diffusion coefficient $D_{\text{pl}}(t)$.
The noise ${\xi_{\text{act}}(t)}$ corresponds to the stress fluctuations produced by activity (cell division and apoptosis), and has a time-independent diffusion coefficient $D_0$.
We assume that these two competing noises are statistically independent.
Hence, the stress diffusion coefficient has two additive contributions,
$D(t) = D_{\text{pl}}(t) + D_0$.
Following the standard H\'ebraud-Lequeux model \cite{hebraud_lequeux_1998_PhysRevLett81_2934}, the diffusion coefficient $D_{\text{pl}}(t)$ modeling the effect of plastic events is self-consistently determined as $D_{\text{pl}}(t) = \alpha \, \Gamma(t)$, where ${\Gamma(t)}$ is the global plastic activity and $\alpha$ is a coupling parameter related to the elastic stress propagator \cite{bocquet_PhysRevLett103_036001,agoritsas_2015_EPJE38_71,SM}.
Here, we emphasize that the new key ingredient is the diffusion coefficient ${D_0>0}$ stemming from activity.
The macroscopic stress $\langle \sigma_{xy} \rangle $ can be obtained from the probability distribution of the local stress $\sigma$.
The evolution equation of this distribution is
\begin{eqnarray}
&& \begin{split}
 \partial_t \mathcal{P}(\sigma,t)
 =	& - G_0 \dot{\gamma} \, \partial_{\sigma} \mathcal{P} +
\argp{\alpha \, \Gamma(t) + D_0} \, \partial_{\sigma}^2 \mathcal{P}
 \\
 -	& \frac{1}{\tau} \theta(\valabs{\sigma}-\sigma_c) \, \mathcal{P}
  + \Gamma(t) \, \delta(\sigma)
\end{split}
\end{eqnarray}
where
${\Gamma(t) = \frac{1}{\tau} \int_{\valabs{\sigma}>\sigma_c} d \sigma \, \mathcal{P}(\sigma,t)}$ is the average number of sites that yield per unit time;
$\Gamma(t)$ is proportional to the number of sites that have reached 
the threshold (i.e., ${\valabs{\sigma_i}>\sigma_c}$) divided by 
the ``lifetime'' $\tau$
\cite{hebraud_lequeux_1998_PhysRevLett81_2934,agoritsas_2015_EPJE38_71}.
The stress on these sites is reset to zero after the yield event. These simple rules do not correspond to the true local relaxation processes \cite{Puosi_Olivier_Martens_2015_Softmatter11_7639}. The aim is to introduce the simplest analytically solvable mean-field scenario that predicts qualitatively well the numerical data with a set of few meaningful effective parameters.

In the steady state at constant shear rate, and in the absence of activity (${D_0=0}$), it is well known that this mean-field model predicts the existence of a Herschel-Bulkley regime with an exponent ${n=1/2}$ at low $\dot{\gamma}$ and for ${\alpha < \sigma_c^2/2}$ \cite{hebraud_lequeux_1998_PhysRevLett81_2934,agoritsas_2015_EPJE38_71,agoritsas_martens_2016_ArXiv-1602.03484}.
%

In the presence of activity ($D_0>0$), the present mean-field model
reproduces the fluidization process leading to a linear behavior (i.e., Newtonian regime) and, quite importantly, recovers a nonlinear flow curve beyond a crossover shear rate $\dot{\gamma}^*$,
\begin{equation}
\label{Eq:3}
\langle \sigma_{xy}\rangle \approx   
\begin{cases}
	\eta \, \dot{\gamma} & \mbox{if } \dot{\gamma} < \dot{\gamma}^* \vspace{0.1cm} \\
	\sigma_y + A_{\text{HB}} \, \dot{\gamma}^{1/2} & \mbox{if } \dot{\gamma} > \dot{\gamma}^*
\end{cases}
\end{equation}
as observed in Fig.~\ref{fig:2}.
Analytical calculations show that $\dot{\gamma}^* \sim D_0$ for $D_0 \to 0$.
The explicit expressions for ${\lbrace \eta, \sigma_y, A_{\text{HB}}, \dot{\gamma}^{*} \rbrace}$ can be computed  as a function of the model parameters ${\lbrace G_0, \tau, \sigma_c, \alpha, D_0 \rbrace}$, using the methods described in Ref.~\cite{agoritsas_2015_EPJE38_71} --~see the Supplemental Material in Ref.~\cite{SM}.


To compare the mean-field model with the numerical data of the particle-based model, we have to fit the values of the model parameters ${\lbrace G_0, \tau, \sigma_c, \alpha, D_0 \rbrace}$.
We used the following fitting procedure.
First, the elastic modulus $G_0$ is estimated independently from the initial elastic response in the stress-strain curve (see Fig.~\ref{fig:1}) yielding ${G_0\approx0.25}$.
Second, the parameters ${\lbrace \tau, \sigma_c, \alpha \rbrace}$ are fitted
on the flow curve obtained in the absence of activity ($a=0$, $\Phi 
\approx 0.94$),
in turn yielding ${\tau =0.12}$, ${\sigma_c=0.15}$, and ${\alpha=0.45 \sigma_c^2}$. Having fixed the four parameters ${\lbrace G_0, \tau, \sigma_c, \alpha \rbrace}$, we then fit the different flow curves obtained in the active case ($a>0$) with $D_0$ as the only free parameter.
The procedure eventually yields the fitted value $D_0(a)$, which is 
plotted in Fig.~\ref{fig:3}(a).
As expected, the fitted value of $\sigma_c$ is larger than 
${\sigma_y}$, see Eq.~\ref{Eq:3},
and that the coupling $\alpha$ is smaller than ${\sigma_c^2/2}$ as required to observe a Herschel-Bulkley behavior. We further observe that the obtained
value of $\alpha/\sigma_c^2$ is larger than for the Lennard-Jones systems \cite{Puosi_Olivier_Martens_2015_Softmatter11_7639};
this larger value might be linked to the presence of a softer potential.
%
\begin{figure}[th]
\begin{center}
\includegraphics[width=0.7\columnwidth, clip]{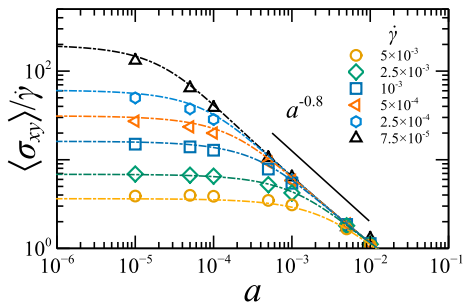}
\end{center}
\caption{
Viscosity ${\eta=\langle \sigma_{xy} \rangle /\dot{\gamma}}$ as a function of the apoptosis rate $a$ for different small values of the shear rate (symbols for the numerical data and dash-dotted lines for the mean-field prediction).
}
\label{fig:4}
\end{figure}
More importantly, the stress diffusion coefficient ${D_0(a)}$ fitted 
from the mean-field prediction is plotted in Fig.~\ref{fig:4}(a), and scales fairly linearly with~$a$.
This can be understood as follows.
At low $a$, apoptosis events are rare and independent, so we can safely assume that the typical redistributed stress ${\Delta \sigma}$ after such events will depend only on the packing fraction but not on the activity.
Moreover, at low shear rates the only relevant time scale is set by the apoptosis rate.
Thus, we can estimate the stress diffusion coefficient $D_0$ by a simple scaling argument as ${D_0\sim (\Delta \sigma)^2 a}$.
Furthermore, the mean-field picture predicts a crossover 
${\dot{\gamma}^*}$ linear with $D_0$, and hence  we expect 
${\dot{\gamma}^*\sim a}$, consistently with the flow curve crossover 
shown in Fig.~\ref{fig:3}(b).

To provide further insights on the behavior of our model,
we show in Fig.~\ref{fig:4} the activity-dependent viscosity ${\eta=\langle \sigma_{xy} \rangle /\dot{\gamma}}$ for different fixed,
low values of the shear rate.
We observe that the viscosity decreases when the activity is enhanced beyond a (strain-rate-dependent) threshold value, i.e., fluidizing the system as previously reported in Ref.~\cite{Ranft2010}.
However, the Maxwell picture proposed in Ref.~\cite{Ranft2010} cannot capture the crossover to the nonlinear rheology that we observe in Fig.~\ref{fig:2}(a).
This is due to the fact that the plateau in the viscosity observed at low activity [see Fig.~\ref{fig:4}] does not correspond to a Newtonian regime
(in which the shear stress varies linearly with $\dot{\gamma}$),
but rather to a yield stress fluid behavior.
This is clearly seen from the fact that the plateau value of the viscosity strongly depends on the strain rate $\dot{\gamma}$, a property that cannot be accounted for by a simple Maxwell model.
The data presented in Fig.~\ref{fig:4} can be understood as follows.
At low enough activity, plasticity is dominant and the shear stress remains independent of activity. For a fixed shear rate, the viscosity ${\eta=\langle \sigma_{xy} \rangle /\dot{\gamma}}$ is thus independent of activity, yielding the plateau in Fig.~\ref{fig:4}.
At a higher activity, the active contribution to the mechanical noise 
instead becomes dominant, thus decreasing the viscosity.
In this regime, the mean-field model predicts 
${\eta \sim 1/D_0}$ (see the Supplemental Material in Ref.~\cite{SM}),
in good agreement with Fig.~\ref{fig:4} (dashed-dot lines), taking into account
$D_0 \sim a$.
Note here again the good agreement between the mean-field model and the numerical simulations.


In conclusion,
we proposed in this Letter a generic scenario to understand the 
crossover from linear to nonlinear rheology in flowing active tissues, based on stress fluctuations mediated by long-range elastic interactions.
The mean-field picture presented here allows us to introduce explicitly the interplay of the two relevant time scales in these systems, one imposed by the external shear and another by the internal processes of the biological tissue in the form of cell division and apoptosis.
This scenario is able to rationalize well the numerical findings of our 
particle-based model for the active confluent tissue under shear, as 
can be seen from the flow curves and the activity-dependent viscosity.

It has been shown that our mean-field predictions are qualitatively robust to the addition of disorder \cite{agoritsas_2015_EPJE38_71}, the partial relaxation of stress or the effective shear-rate dependence of the elastic modulus or relaxation time \cite{agoritsas_martens_2016_ArXiv-1602.03484}, assessing furthermore the generality of our scenario. 
Moreover, the introduction of additional relaxation mechanisms like cell-shape fluctuations, self propulsion, external vibrations, and other sources of mechanical noise, can be easily implemented in our mean-field description via assumptions on the distribution of the active part of the noise.
However, to allow for a more refined description of the dynamics aiming 
for a quantitative agreement, it would be interesting to investigate in 
more details the long-range elastic effects of cell division and 
apoptosis events. 
A strong point of our approach is that it can be easily generalized to 
describe the rheological response of other systems that include an additional shear-rate independent noise, such as vibrated grains \cite{d2003observing, ford2009transitions, dijksman_2011_PhysRevLett107_108303, Pons2016}, active colloidal suspensions \cite{Theurkauff2012} or coarsening foams \cite{hilgenfeldt2001dynamics, Saint-Jalmes2006}.


\begin{acknowledgments}
\paragraph{Acknowledgements.}
J.-L.~B., D.~A.~M.-F. and E.~A.~acknowledge financial support from European Research Council Grant No.~ADG20110209; E.~A.~was also supported by the Swiss National Science Foundation Grant No P2GEP2-15586 and by a Simons Foundation Grant ($\sharp$ 454955, Zamponi).
J.-L.~B., D.~A.~M.-F., and K.~M. thank the NVIDIA Corporation for a hardware grant through the Academic Partnership Program.
Further, we would like to thank Silke Henkes and Rastko Sknepnek for valuable discussions on the particle model description during this work.

D.~A.~M.-F. and E.~A. contributed equally to this work.
\end{acknowledgments}



%

\end{document}